\documentclass[11pt]{article}
\usepackage{mathrsfs}
\usepackage{amsmath}
\usepackage{hyperref}
\usepackage{amsfonts}
\usepackage{amssymb, amsmath, cite}
\usepackage{color}
\usepackage{graphicx}
\usepackage{array}
\usepackage{booktabs}
\usepackage{makecell}
\usepackage{diagbox}
\usepackage[T1]{fontenc}
\usepackage{tabularx}
\usepackage{amsthm}

\usepackage{pgfplots}

\pgfplotsset{compat=1.18}

\numberwithin{equation}{section}
\numberwithin{figure}{section}

\newtheorem{theorem}{Theorem}[section]

\newcommand\be{\begin{equation}}
\newcommand\ee{\end{equation}}
\newcommand\ber{\begin{eqnarray}}
\newcommand\eer{\end{eqnarray}}
\newcommand\berr{\begin{eqnarray*}}
\newcommand\eerr{\end{eqnarray*}}
\newcommand\bea{\begin{eqnarray}}
\newcommand\eea{\end{eqnarray}}

\newcommand{\bfR}{{\Bbb R}}

\newcommand{\dd}{\mbox{d}}
\newcommand{\e}{\mbox{e}}
\newcommand{\pa}{\partial}

\newcommand{\nn}{\nonumber}

\newcommand\lb{\label}
\newcommand\eq{\eqref}

\title{Universal Properties of Nonlinearly Perturbed Maxwell Theory}

\author{{Tengyang Liu}$^{a,}$\footnote{Email address: tyl619@henu.edu.cn} { and Yisong Yang}$^{b,}$\footnote{Email address: yisongyang@nyu.edu}\\[2mm]{\it\small $^a$College of Mathematics and Statistics, Henan University}\\[2mm]{\it\small Kaifeng 475001, P.R.China}\\[2mm]{\it\small $^b$Courant Institute of Mathematical Sciences}\\{\it\small New York University}\\{\it\small New York, New York 10012, USA}}
\date{}

\begin{document}

\maketitle

\begin{abstract}
We show that a general nonlinearly perturbed Maxwell theory of electromagnetism possesses three universal fundamental properties: 
\begin{itemize}
\item A finite-energy electric point charge

\item Exclusion of finite-energy magnetic monopoles and dually charged dyons

\item Arbitrary smallness of the effective radius of a point electric charge and the associated local undetectedness of the electric charge and energy

\end{itemize}

In particular, this last property offers a classical explanation for the invisibility of the electron, as a point electric charge, in accordance with the smallness of its effective radius.

This nonlinear theory of electromagnetism has the feature that it minimally perturbs the Maxwell theory with a nonlinearity profile that is
as general as possible such that the three universal properties stated above are all maintained.

\medskip

\medskip




\medskip

\medskip

\medskip

\end{abstract}

\section{Introduction}
\setcounter{equation}{0}

It is well known that the Born--Infeld theory \cite{B1,B2,BI1,BI2} was introduced as a nonlinear extension of the linear Maxwell theory to accommodate a {\em finite-energy} electric point charge
in order to model the electron as a point particle, which is impossible in the Maxwell theory. Despite its success in achieving this purpose, it is also limited by the fact that in all its parameter
regime the effective radius for the electron it produces is of the order $10^{-13}$ cm, within the reach about 60--70\% of the charge and mass are maintained, which is much bigger than
the laboratory established electron radius bound \cite{Deh1,Deh2,Deh3} under $10^{-20}$ cm. In fact, the electron as a classical particle has been undetectable as of today. Besides, the classical Born--Infeld theory also
accommodates a finite-energy magnetic point charge known as the monopole mathematically conceptualized first by Dirac \cite{Dirac} in the Maxwell theory. In theoretical physics, the existence of monopoles is a natural consequence of the non-Abelian gauge field
theory structure but has remained a puzzle as an electromagnetic construct:  Like an electric point charge, a monopole carries a finite or infinite energy in the Born--Infeld theory or
Maxwell theory, respectively. In other words, in both Maxwell and Born--Infeld theories, monopoles and electric point charges  are equally accommodated energetically. Since the monopole
as a magnetic point charge has never been observed in laboratory, it will be of fundamental interest to develop a theory of electromagnetism in which finite-energy electric point charges
are allowed but magnetic point charges will inevitably encounter energy divergence, such that monopoles are excluded energetically. In \cite{LY1}, we developed a theory that
minimally interpolates
the Maxwell theory and the Born--Infeld theory defined by the Lagrangian action density
\be\lb{a1.1}
{\cal L}=-\frac{\kappa_1}4F_{\mu\nu}F^{\mu\nu}+\frac{\kappa_2}\beta\left(1-\sqrt{1-2\beta s}\right),
\ee
where $F_{\mu\nu}=\partial_{\mu}A_\nu-\partial_\nu A_\mu$ is the electrodynamics field tensor induced by a real-valued gauge field $A_{\mu}$, $\beta=b^{-2}$ with $b>0$ being
the Born parameter, $\kappa_1,\kappa_2>0$ are dimensionless interpolation parameters satisfying $\kappa_1+\kappa_2=1$,
\begin{align}
s=-\frac{1}{4}F_{\mu\nu}F^{\mu\nu}+\frac{\kappa^2}{32}\left(F_{\mu\nu}\widetilde{F}^{\mu\nu} \right)^2,\quad \widetilde{F}^{\mu\nu}=\frac{1}{2}\epsilon^{\mu\nu\alpha\beta}F_{\alpha\beta},\label{a1.2}
\end{align}
with $\kappa\geq 0$ being a free coupling parameter, and $\mu,\nu$ the Minkowski spacetime coordinate indices with the metric $g_{\mu\nu}=\mbox{diag}(1,-1,-1,-1)$. We showed that this
theory accommodates finite-energy electric point charges as in the Born--Infeld theory but unlike the Born--Infeld theory it excludes monopoles in all ranges of the parameters. Furthermore,
the effective electron radius of the theory remains to be in the order of $10^{-13}$ cm as that in the Born--Infeld theory. That is, a small enough effective radius of the electron is 
still unreachable
by the model \eq{a1.1}.

Extending the spirit of the model \eq{a1.1}, a perturbed Maxwell--Born--Infeld model that couples a pure Maxwell term with a  Born--Infeld type nonlinearity was introduced in \cite{LY2} with the Lagrangian action density
\begin{align}
\mathcal{L}=-\frac14F_{\mu\nu}F^{\mu\nu}+\frac{\gamma}{\beta}\Bigl(1-\sqrt{1-(2\beta s)^k}\Bigr),\quad  k=3,5,\dots.\label{1.1}
\end{align}
 It is shown \cite{LY2} that the effective electron radius of the model \eq{1.1} can be made far below the experimental limit when the coupling parameter $\gamma$
is chosen to be sufficiently large. Consistently, it is also demonstrated \cite{LY2} that, in the large $\gamma$ or zero effective radius limit, the free electric charge and self energy contained within any ball
around the point charge tend to zero, which provides a classical field-theoretical interpretation of the invisibility of the electron as a point particle. Besides, like the interpolation model \eq{a1.1},
the perturbation model \eq{1.1} enjoys the exclusion of  monopoles as a result of the presence of the Maxwell term.

An intriguing feature, perhaps surprising and interesting too, of the model \eq{1.1} is that the self energy of an electric charge stays finite in the limit $\gamma\to 0$. In other words, the model
does not return to the Maxwell theory in its Maxwellian limit. This puzzle leads us to consider the polynomially perturbed model
\begin{align}
\mathcal{L}=-\frac14F_{\mu\nu}F^{\mu\nu}+\frac{\gamma}\beta \sum_{k=2}^n b_k (\beta s)^k,\quad b_k\geq0,\quad k=2,\dots,n,\quad b_n>0,\label{a1.4}
\end{align}
in \cite{LY2}. It is shown that the self energy of an electric charge blows up as $\gamma\to0$, thereby recovering the Maxwell theory limit. Moreover, this model maintains all other desired
features of the model \eqref{1.1}, including accommodating finite-energy electric point charges, exclusion of monopoles, arbitrary smallness of the effective radius and local undetectability of
an electric point charge in the large $\gamma$ limit.

These common and different features exhibited by the models \eq{1.1} and \eq{a1.4} suggest that their shared structures and distinctions should be explored in a unified
way in the large picture of the theory of nonlinear electrodynamics in order to understand from where and how these features arise.
In fact, two clearly shared structures of \eq{1.1} and \eq{a1.4} are that they both contain the Maxwell density as the leading term in their action densities and the nonlinear terms
are of higher-order with an adjustable coupling parameter $\gamma$ as a modulator for phenomenological fine tuning. On the other hand, a sharp distinction of these models is that 
\eq{1.1} imposes a truncation threshold $2\beta s\leq 1$ but \eq{a1.4} does not. These shared and distinct structures and the associated universal properties anticipated from and implicated by
these structures are the main interests
of our study here.

Specifically, in this work, we shall consider a general nonlinearly perturbed Maxwell theory governed by the Lagrangian action density
\begin{align}
\mathcal{L}=-\frac14F_{\mu\nu}F^{\mu\nu}+\frac{\gamma}{\beta}U(\beta s),
\label{1.5}
\end{align}
where $U$ is a nonlinear perturbation term satisfying the condition $U(0)=0, U'(0)=0$ such that the nonlinearity it introduces is of higher order. The main universal properties set as our 
goal to obtain,
as already stated, are formally reformulated and articulated as follows:
\begin{enumerate}
\item[(i)] The accommodation of finite-energy electric point charges.

\item[(ii)] The exclusion of finite-energy magnetic monopoles and dually charged point particles known as dyons.

\item[(iii)] The capability of achieving arbitrary smallness of the effective radius of a point electric charge and the associated local undetectedness of the charge, either electrically or
electrically {\em and} energetically.
\end{enumerate}

We will show that these properties are indeed the universal properties of the general model \eq{1.5}.

As discussed regarding the models \eq{1.1} and \eq{a1.4}, the structure of \eq{1.5} leads to the following intuitive anticipation regarding an electric point charge:
\begin{enumerate}
\item[(iv)] The self energy of an electric point charge tends to infinity as $\gamma\to0$ thereby recovering the Maxwell theory limit known as the ultraviolet divergence.

\end{enumerate}

We will show that whether this property holds depends on the absence or presence of a finite truncation threshold as exhibited in the models \eq{a1.4} and \eq{1.1}, respectively. 
As a consequence, this study classifies the general model \eq{1.5} into two categories: one that possesses the property (iv) and one that does not. In the latter situation, we show
that the self energy of an electric point charge approaches a finite limit as $\gamma\to0$ as that in the model \eq{1.1}. Since this is also a valid property for the general model \eq{1.5}
when a truncation threshold exists, it may be regarded as a universal property as well and stated as follows:

\begin{enumerate}
\item[(v)] The self energy of an electric point charge tends to a finite value as $\gamma\to0$ such that the Maxwell theory limit is not unreachable.

\end{enumerate}

In subsequent sections, we carry out our work. In the last section, we summarize our results obtained and make some further comments.

\section{Model formulation and exclusion of monopoles and dyons}\label{s2}
\setcounter{equation}{0}

Varying the gauge field $A_\mu$ in \eq{1.5}, we get the Euler--Lagrange equations
\be\lb{2.1}
\pa_\mu P^{\mu\nu}=j^\nu,
\ee
where $j^\nu$ is an external 4-current density and
\be\lb{2.2}
P^{\mu\nu}=F^{\mu\nu}+\gamma U'(\beta s)\left(F^{\mu\nu}-\frac{\kappa^2}4(F_{\mu'\nu'}\tilde{F}^{\mu'\nu'})\tilde{F}^{\mu\nu}\right)
\ee
generates the electric displacement field ${\bf D}=(D^i)$ and the magnetic intensity field ${\bf H}=(H^i)$ through the realization 
\begin{align}\lb{2.3}
    (P^{\mu\nu})=\left(\begin{matrix}
        0&-D^1&-D^2&-D^3\\
       D^1&0&-H^3&H^2\\
      D^2&H^3&0&-H^1\\
        D^3&-H^2&H^1&0
    \end{matrix}\right).
\end{align}
Besides,
using ${\bf E}=(E^i)$ and ${\bf B}=(B^i)$ to denote the electric and magnetic fields, respectively, we have their tensor matrix representations
\begin{align}
    (F_{\mu\nu})=\left(\begin{matrix}
        0&E^1&E^2&E^3\\
       - E^1&0&-B^3&B^2\\
      -  E^2&B^3&0&-B^1\\
        -E^3&-B^2&B^1&0
    \end{matrix}\right),\quad
(\tilde{F}^{\mu\nu})=\left(\begin{matrix}
        0&-B^1&-B^2&-B^3\\
       B^1&0&E^3&-E^2\\
      B^2&-E^3&0&E^1\\
        B^3&E^2&-E^1&0
    \end{matrix}\right).
\label{2.4}
\end{align}
Thus the Maxwell action density and the Born--Infeld scalar $s$ given in \eq{a1.2} read
\bea
-\frac{1}{4}F_{\mu\nu}F^{\mu\nu}&=&\frac12({\bf E}^2-{\bf B}^2),\\
s&=&
\frac{1}{2}(\textbf{E}^2-\textbf{B}^2)+\frac{\kappa^2}2(\textbf{E}\cdot\textbf{B})^2,\label{1.3a}
\eea
and \eq{2.2} leads to the constitutive equations
\bea
{\bf D}&=&{\bf E}+\gamma U'(\beta s)\left({\bf E}+\kappa^2({\bf E}\cdot{\bf B}){\bf B}\right),\lb{2.7}\\
{\bf H}&=&{\bf B}+\gamma U'(\beta s)\left({\bf B}-\kappa^2({\bf E}\cdot{\bf B}){\bf E}\right).\lb{2.8}
\eea
Thus, using $j^\nu=(\rho_e,{\bf j}_e)$ to denote the electric charge and current densities, we can insert \eq{2.3} and \eq{2.4} into \eq{2.1} to obtain the static covariant Maxwell equations
\bea
\nabla\cdot{\bf D}&=&\nabla\cdot\left({\bf E}+\gamma U'(\beta s)\left({\bf E}+\kappa^2({\bf E}\cdot{\bf B}){\bf B}\right)\right)=\rho_e,\lb{2.9}\\
\nabla\times{\bf H}&=&\nabla\times \left({\bf B}+\gamma U'(\beta s)\left({\bf B}-\kappa^2({\bf E}\cdot{\bf B}){\bf E}\right)\right)={\bf j}_e,\lb{2.10}\\
\nabla\cdot{\bf B}&=&\rho_m,\lb{2.11}\\
\nabla\times{\bf E}&=&-{\bf j}_m,\lb{2.12}
\eea
following the formalism of Schwinger \cite{Sch1,Sch2,Sch3} with {\em imposed} magnetic charge and current densities $\rho_m$ and ${\bf j}_m$ in order to restore electromagnetic duality.

Furthermore, formally varying the metric tensor $g_{\mu\nu}$ in \eq{1.5}, we obtain
the energy-momentum tensors
\begin{align}
    T_{\mu\nu}=-F_{\mu\mu'}g^{\mu'\nu'}F_{\nu\nu'}
   -\gamma U'(\beta s)\left(F_{\mu\mu'}g^{\mu'\nu'}F_{\nu\nu'}-\frac{\kappa^2}{4}\left(F_{\mu'\nu'}\tilde{F}^{\mu'\nu'}\right)F_{\mu\mu''}g^{\mu''\nu''}\tilde{F}_{\nu\nu''}\right)-g_{\mu\nu}{\cal L}.
\end{align}
Thus, inserting \eq{2.4}, we see that the associated Hamiltonian energy density reads
\begin{align}
\mathcal{H}=T_{00}=\frac{1}{2}\left(\textbf{E}^2+\textbf{B}^2\right)+\gamma\left((\textbf{E}^2+\kappa^2(\textbf{E}\cdot\textbf{B})^2)U'(\beta s)-\frac{1}{\beta}U(\beta s)\right). 
\label{2.14}
\end{align}

We now consider a solution to \eq{2.9}--\eq{2.12} with a nontrivial magnetic charge distribution which is point-like at some ${\bf x}_0\in{\mathbb R}^3$. Then we have
\be
\rho_m({\bf x})=4\pi q_m\delta({\bf x}-{\bf x}_0),\quad {\bf x}\in{\mathbb R}^3,\quad r=|{\bf x}-{\bf x}_0|\ll1,
\ee
where $q_m\neq0$ is a local magnetic charge and $\delta({\bf x})$ is the Dirac distribution concentrated at the origin. Solving \eq{2.11} near ${\bf x}_0$, we have
\be
{\bf B}({\bf x})=\frac{q_m({\bf x}-{\bf x}_0)}{r^3},\quad {\bf x}\neq {\bf x}_0,\quad r=|{\bf x}-{\bf x}_0|\ll1.\label{2.16}
\ee
In view of this expression,
we see that the self energy calculated through integrating \eq{2.14} over $\bfR^3$ inevitably diverges. 

In summary, we state the following monopole and dyon exclusion principle:

\begin{theorem}\lb{th2.1}

In all the parameter regime, the fundamental governing equations \eq{2.9}--\eq{2.12} of the nonlinearly perturbed Maxwell electromagnetic theory defined by the general action density \eq{1.5} do not accommodate any finite-energy locally point-like solutions with
a nontrivial magnetic charge density. In particular, finite-energy monopoles and dyons are excluded in this theory.

\end{theorem}

\medskip

The significance of Theorem \ref{th2.1} lies in its rigorous demonstration that the non-Maxwellian perturbations considered here do not merely modify the field profiles, but fundamentally restrict the admissible topological sectors of the theory. In standard Maxwell electrodynamics, magnetic monopoles are excluded only by the ad hoc assumption of a vanishing magnetic charge density in the Bianchi identity. Here, however, the exclusion here is a consequence of the energetic consistency within the nonlinear framework: The presence of a nontrivial magnetic or dyonic 
point source would necessitate a configuration whose total energy is non-integrable. This result effectively establishes that the ``universal" properties of this perturbed theory are inherently ``purely electric." It further suggests that any pursuit of magnetic singularities within this class of Lagrangians would require a departure from the finite-energy requirement.

In the next section, we will consider purely electrically charged solutions to the system \eq{2.9}--\eq{2.12}.

\section{Electric charge problem in a general setting}\label{s3}
\setcounter{equation}{0}

In view of Theorem \ref{th2.1},  we see that only purely electrically charge solutions are possible for the system \eq{2.9}--\eq{2.12} for locally point-like field configurations under the
finite energy condition. In this situation, the magnetic field $\bf B$ and magnetic intensity field $\bf H$ are trivial and the constitutive equations \eq{2.7}--\eq{2.8} are reduced into a single one:
\be\lb{a3.1}
{\bf D}=(1+\gamma U'(\beta s)){\bf E},\quad s=\frac12{\bf E}^2.
\ee
From \eq{a3.1}, we have
\be\lb{3.2}
\beta{\bf D}^2=2(1+\gamma U'(\beta s))^2\beta s\equiv g(\eta),\quad \eta\equiv \beta{\bf E}^2.
\ee
Formally we assume that \eq{3.2} can be inverted to yield the solution
\be\lb{a3.3}
\beta{\bf E}^2=\eta=h(\beta{\bf D}^2),
\ee
where $h=g^{-1}$. In order to ensure that this procedure can be carried out, it suffices to assume the condition $g'(\eta)>0$ or
\be\lb{3.4}
(1+\gamma U'(\tau))(1+\gamma U'(\tau)+2\gamma\tau U''(\tau))>0,\quad \tau=\beta s,
\ee
in the domain of interest, which in turn is ensured by imposing the condition
\be\lb{b3.5}
U'(\tau)\geq0,\quad 1+\gamma U'(\tau)+2\gamma\tau U''(\tau)>0,\quad \tau\geq0,
\ee
for convenience, which will be assumed from now on.
Inserting \eq{a3.3} back into \eq{a3.1}, we obtain
\be\lb{a3.5}
{\bf E}=\frac{{\bf D}}{1+\gamma U'\left(\frac12h(\beta{\bf D}^2)\right)}.
\ee

We now examine some basic properties contained in the inverted electric constitutive equation \eq{a3.5}. 

First, from the Poisson equation \eq{2.9}, we have the total electric charge
\bea\lb{a3.6}
Q&=&\int_{\bfR^3} \rho_e({\bf x})\,\dd{\bf x}\nn\\
&=&\lim_{R\to\infty}\int_{|{\bf x}|=R}{\bf D}\cdot \dd {\bf S}.
\eea
Thus, assuming the finite-charge condition, we arrive at the asymptotic property
\be\lb{3.7}
{\bf D}({\bf x})=\mbox{O}(r^{-2}),\quad r=|{\bf x}|\gg1,
\ee
which is simply the classical Coulomb law. Using \eq{3.7} in \eq{a3.5}, we see that $\bf E$ is also of  the Coulomb type. That is, it satisfies the same asymptotic property as $\bf D$.

Next, following the formalism of Born and Infeld \cite{BI1,BI2},  we define the free electric charge density by the electric field $\bf E$ by
\be\lb{a3.8}
\rho_{\rm{free}}=\nabla\cdot{\bf E}.
\ee
Hence the total free electric charge may be calculated through the divergence theorem by
\bea
Q_{\rm{free}}&=&\int_{\bfR^3}\rho_{\rm{free}}({\bf x})\,\dd{\bf x}\nn\\
&=&\lim_{R\to\infty}\left(\int_{|{\bf x}|=R}{\bf D}\cdot\dd{\bf S}+\int_{|{\bf x}|=R}\left(\frac{{\bf D}}{1+\gamma U'\left(\frac12h(\beta{\bf D}^2)\right)}-{\bf D}\right)\cdot\dd{\bf S}\right)\nn\\
&=&Q,\label{a3.9}
\eea
 using \eq{a3.6}, \eq{3.7}, and the condition $U'(0)=0$.

With the electric field $\bf E$ given in \eq{a3.5}, we can compute the magnetic current ${\bf j}_m$ whose presence is necessary for the balance of a multicentered electric charge system.
However, if the system is singly centered, ${\bf j}_m$ vanishes as established in \cite{Y-AOP}.

Finally, in the electrostatic situation under consideration, the Hamiltonian energy density \eq{2.14} becomes
\be\lb{a3.10}
{\cal H}=\frac12{\bf E}^2+\gamma \left({\bf E}^2 U'(\beta s)-\frac1\beta U(\beta s)\right).
\ee
Hence the self energy always converges near infinity of the space as a consequence that $\bf E$ satisfies the same decay law \eq{3.7} as the electric displacement field
$\bf D$. On the other hand, if the electric charge is localized like a point charge $q\neq 0$ concentrated at the origin,  say,  then ${\bf D}$ has the expression
\be\lb{a3.11}
{\bf D}({\bf x})=\frac{q{\bf x}}{r^3},\quad {\bf x}\neq{\bf 0},\quad r=|{\bf x}|\ll1.
\ee
Thus, we may express the energy density \eq{a3.10} of an electrostatic system as
\be\lb{a3.12}
{\cal H}=\left(\frac12+\gamma U'\left(\frac12h(\beta{\bf D}^2)\right)\right)\frac{{\bf D}^2}{\left(1+\gamma U'\left(\frac12h(\beta{\bf D}^2)\right)\right)^2}-\frac\gamma\beta U\left(\frac12h(\beta{\bf D}^2)\right),
\ee
in view of  \eq{a3.5}. Note that we shall always implicitly observe that the relevant nonnegativity condition ${\cal H}\geq0$ in our study.

To proceed further, we convert \eq{3.2} into 
\be\lb{x3.14}
1+\gamma U'\left(\frac\eta2\right)=\sqrt{\frac\beta\eta} |{\bf D}|,\quad \eta=h(\beta{\bf D}^2).
\ee
Inserting \eq{x3.14} into \eq{a3.12}, we get 
\be\lb{x3.15}
{\cal H}=\frac1\beta\left(\sqrt{\beta\eta}|{\bf D}|-\frac\eta2-\gamma U\left(\frac\eta2\right)\right)\leq\sqrt{\frac\eta\beta}|{\bf D}|.
\ee

In the next section, we shall see that the nonlinearity $U$ gives rise to two families of the models, characterized by the properties that $\eta\to\eta_0>0$ as $|{\bf D}|\to\infty$ such that
$U'(\tau)\to\infty$ as $\tau\to \frac{\eta_0}2$, referred to as the finite-truncation case, and $\eta\to\infty$ as $|{\bf D}|\to\infty$, referred to as the no-finite-truncation case, respectively.

In the former situation, using \eq{x3.15}, we see that ${\cal H}=\mbox{O}(r^{-2})$ near a point charge at the origin. As a consequence, we have 
established the convergence of the total energy in this situation as a general result.

In the latter situation, we consider two typical subcases for simplicity and applicability.

\begin{enumerate}

\item[(i)] The function $U$ is dominated by a power function, say $U(\tau)\sim \tau^k$ ($k>\frac32$) for $\tau\gg1$. Then \eq{3.2} gives us $\eta\sim r^{-\frac4{2k-1}}$ near $r=0$, which gives us
${\cal H}\sim r^{-2-\frac2{2k-1}}$ near $r=0$ such that the convergence of the energy follows since $\frac2{2k-1}<1$.

\item[(ii)] The function $U$ is dominated by an exponential function, say $U(\tau)\sim \e^{\tau^k}$ ($k>\frac12$) for $\tau\gg1$. Then \eq{3.2} leads to the relation
\be\lb{3.19}
r^{-2}\sim \tau^{k-\frac12}\e^{\tau^k}\geq W\e^W,\quad W=\tau^{k-\frac12},\quad r\ll1\quad\mbox{or}\quad \tau\gg1.
\ee
Recall that the relation $W\e^W=x$ defines the Lambert $W$-function for $x\geq-\frac1\e$ which increases with respect to $x$ such that
\be\lb{3.20}
W(x)=\ln x-\ln\ln x+\cdots,\quad x\gg1.
\ee
Using \eq{3.20} in \eq{3.19}, we have the bound
\be\lb{3.18}
\tau\leq W^{\frac2{2k-1}}(r^{-2})\leq\left(2\ln\left(\frac1r\right)\right)^{\frac2{2k-1}},\quad r\ll1,
\ee
up to some irrelevant constants. Inserting \eq{3.18} into \eq{x3.15}, we have
\be
{\cal H}\sim r^{-2}\left(\ln\left(\frac1r\right)\right)^{\frac1{2k-1}},\quad r\ll1,
\ee
which leads to the convergence of the self energy again.

\end{enumerate}

Thus our study of this section may be summarized as follows.

\begin{theorem}\lb{th3.1}

In all parameter regime such that the condition \eq{b3.5} is valid, the static covariant Maxwell equations \eq{2.9}--\eq{2.12} have a unique electrostatic
solution realizing an arbitrary electric charge density distribution $\rho_e ({\bf x})$, with the total electric charge $Q$ given in \eq{a3.6}, which may be discrete or continuous or both, characterized by the absence of magnetism
with ${\bf B}={\bf 0}, {\bf H}={\bf 0}, \rho_m=0, {\bf j}_e={\bf0}$. The total free electric charge generated from the free charge density $\rho_{\rm{free}}({\bf x})$ defined by the electric field $\bf E$ as stated  in \eq{a3.8} coincides with the total
prescribed electric charge of the system.
 In addition, if the electric charge is localized as a point charge as described by \eq{a3.11}, then the solution is also
of a finite energy when the model possesses a finite truncation threshold for the electric field as in the Born--Infeld theory or
when the model does not possess a finite truncation threshold but the nonlinearity function $U$ is subject to some general growth conditions such as those in the forms of the two scenarios (i) and (ii) stated above. 
\end{theorem}

\medskip

This theorem establishes that the regularization of the classical Maxwellian singularity is a universal feature inherent to the entire class of nonlinearly perturbed theories considered here. While the standard linear theory inevitably leads to a divergent self energy for a point source, the estimates derived above demonstrate that any perturbation $U$ subject to the 
generally imposed conditions is sufficient to ensure a finite energy configuration. This result confirms that the accommodation of a physically consistent electric point charge does not depend on a specific choice of nonlinear Lagrangian, such as the square-root form of the Born--Infeld theory, but is instead a structural consequence of the nonlinear interaction. Mathematically, the theorem underscores a broad stability in the energy functional: The high-order terms effectively ``tame" the $r^{-4}$ divergence of the energy density near the origin without disrupting the Maxwellian behavior at large distances. Thus, the finite-energy property is promoted from a model-specific success to a fundamental characteristic of the generalized nonlinear electromagnetic framework.

With this general study, in the next section, we shall focus on the point charge problem.

\section{The electric point charge problem}

\setcounter{equation}{0}

Consider an electric point charge placed at the space origin with charge $q>0$ such that the associated electric displacement field $\bf D$ is given by  \eq{a3.11}
for ${\bf x}\neq{\bf 0}$. Hence $\bf D$ is radially 
symmetric and in terms of spherical coordinates the only nontrivial component of $\bf D$ is in its radial direction, $D_r =\frac{q}{r^2}$.
Thus the induced electric field $\bf E$ is also radial  ${\bf{E}}=(E_r,0,0)$. Inserting these into \eq{a3.1}, we get
\be\lb{x4.1}
\frac{q}{r^2}=(1+\gamma U'(\beta s))E_r,\quad s=\frac12 E_r^2,\quad r>0,
\ee
which gives us the relation
\begin{align}
 \frac{a^4}{r^4}=   \frac{\beta q^2}{r^4}=\left(1+\gamma U'\left(\frac\eta2\right)\right)^2\eta,\quad a=\beta^{\frac14}q^{\frac12},\quad \eta=2\beta s=\beta E_r^2,\label{3.1}
\end{align}
where $a$ is known as the effective radius, which makes this relation dimensionless, and clearly appears as a scale parameter.
Imposing consistency in \eq{3.1}, we arrive at the boundary conditions
\begin{align}
    r\to0,\quad\eta\to\eta_0;\quad r\to\infty,\quad\eta\to0,\label{3.3}
\end{align}
where $\eta_0$ designates the normalized (dimensionless) limiting electric field strength near the charge center which is finite if and only if the nonlinearity function $U$ contains a finite truncation
threshold $\eta_0>0$ realized by
\be\lb{x4.4}
U'\left(\frac\eta2\right)\to\infty,\quad \eta\to\eta_0,
\ee
 otherwise we understand $\eta_0=\infty$. These two scenarios are realized by the radical root type model and polynomial model, respectively, in \cite{LY2}. Here these two scenarios will be given a
unified treatment in our general setting.

In terms of the variable $\eta$ defined in \eq{3.1}, we rewrite the Hamiltonian energy density \eq{a3.10} as
\begin{align}
    {\cal H}=&\frac1\beta\left(\frac{\eta}{2}+\gamma\left(\eta U'\left(\frac\eta2\right)-U\left(\frac\eta2\right)\right)\right).\label{3.5}
\end{align}
Observing the nonnegativity condition ${\cal H}\geq0$, we have
\be\lb{4.6x}
\tau+\gamma(2\tau U'(\tau)-U(\tau))\geq 0,\quad \tau\geq0.
\ee

We note that the integrand of electric self energy 
\be\lb{4.7}
E=4\pi\int^\infty_0\mathcal{H}r^2{\rm{d}}r
\ee
 enjoys a double monotonicity property in terms of $r$ as seen in
\begin{align}
\frac{{\rm{d}}\mathcal{H}}{{\rm{d}}r}&=-\frac{2q^2}{r^5(1+\gamma U'(\frac\eta2))},\label{10a}\\
\frac{{\rm{d}}(\mathcal{H}r^2)}{{\rm{d}}r}&=-\frac{r}\beta\left(\eta+2\gamma U\left(\frac\eta2\right)\right),\label{3.6}
\end{align}
after using the derivatives
\begin{align}
\frac{{\rm{d}}\mathcal{H}}{{\rm{d}}\eta}=&\frac1{2\beta}\left(1+\gamma\left(U'\left(\frac\eta2\right)+\eta U^{\prime\prime}\left(\frac\eta2\right)\right)\right),\label{4.6}\\
\frac{{\rm{d}}\eta}{{\rm{d}}r}=&\frac{-4a^4}{r^5(1+\gamma U'(\frac\eta2))\left(1+\gamma\left(U'\left(\frac\eta2\right)+\eta U^{\prime\prime}\left(\frac\eta2\right)\right)\right)},\label{4.11}
\end{align}
and \eq{3.1}, and going through some neat cancellations.

Recall from \eq{3.1} that $\eta=\beta E_r^2$. Thus, physically, the energy density $\cal H$ is expected to be increasing with respect to $\eta$. In fact, this expectation 
is validated in view of \eq{4.6} and the imposed condition \eq{b3.5}.

Inserting \eqref{3.3}, \eqref{3.5}, and \eq{4.11} into \eq{4.7}, we obtain
\begin{align}
    \frac{E}{4\pi}
=&\frac{q^2}{a}\int^{\eta_0}_0\frac{\left(\frac\eta2+\gamma(\eta U'-U)\right)\left(1+\gamma(U'+\eta U^{\prime\prime})\right)}{4\eta^\frac74(1+\gamma U^{\prime})^\frac52}{\rm{d}}\eta\notag\\
    \equiv&\frac{q^2}{a}\int^{\eta_0}_0h_\gamma(\eta){\rm{d}}\eta
    =\left(\frac{q^2}{a}\right)H_\gamma,\label{3.13}
\end{align}
where $H_\gamma$ is a normalized dimensionless energy and 
\be\lb{4.14}
h_{\gamma}(\eta)=\frac{\left(\frac\eta2+\gamma\left(\eta U'\left(\frac\eta2\right)-U\left(\frac\eta2\right)\right)\right)\left(1+\gamma\left(U'\left(\frac\eta2\right)+\eta U^{\prime\prime}\left(\frac\eta2\right)\right)\right)}{4\eta^\frac74\left(1+\gamma U^{\prime}\left(\frac\eta2\right)\right)^\frac52}\geq0,
\ee
in view of the conditions \eq{b3.5} and \eq{4.6x}.
Therefore, using Fatou's lemma, we have
\begin{align}
    \liminf_{\gamma\to0}H_\gamma\ge\int^{\eta_0}_0 \liminf_{\gamma\to0}h_\gamma(\eta){\rm{d}}\eta=\int^{\eta_0}_0\frac{{\rm{d}}\eta}{8\eta^\frac34},\label{3.16}
\end{align}
which diverges when $\eta_0=\infty$. On the other hand, when $\eta_0<\infty$, we have the direct conclusion
\be
\lim_{\gamma\to0} H_\gamma=\lim_{\gamma\to0}\int_0^{\eta_0}h_\gamma(\eta)\,\dd\eta=\int_0^{\eta_0}\frac{\dd\eta}{8\eta^{\frac34}}=\frac{\eta_0^{\frac14}}{2}.
\ee

The strong-perturbation limit $\gamma\to\infty$ is more subtle. The expression \eq{4.14} clearly gives us the pointwise convergence
\be\lb{4.17}
\lim_{\gamma\to\infty} h_\gamma(\eta)=0.
\ee
Hence, when $\eta_0$ is finite, we have
\be\lb{4.18}
\lim_{\gamma\to\infty}H_\gamma=0.
\ee

To achieve the same consistent limit when $\eta_0=\infty$, it suffices to impose the condition
\be\lb{4.19}
h_\gamma(\eta)\leq \varphi(\eta),\quad \gamma\gg1,
\ee
 where $\varphi(\eta)$ is in $L^1(0,\infty)$, such that \eq{4.18} follows from \eq{4.17} and \eq{4.19} by virtue of the dominated convergence theorem. In view of the structure of \eq{4.14}, 
it is seen that \eq{4.19} is quite a general condition for a broad range of nonlinearity functions $U$ (for example, it is readily satisfied under the two growth scenarios (i) and (ii)
given in Section \ref{s3}). In other words,  the property \eq{4.18} is a general realizable property when $\eta_0=\infty$
as well. In what follows, we shall not elaborate on the technical issues regarding how to ensure \eq{4.19} but consider the consequence of \eq{4.18}.

Consider an electric point charge $q$ placed at the origin which carries a self energy $\frac{E}{4\pi}$ often represented by its rest mass, say $m$.  In view of  \eq{3.13}, we have
\be\lb{4.20}
a=\frac{q^2}{m}H_\gamma.
\ee
This expression leads us to conclude with the following theorem.

\begin{theorem} \label{th4.1}
For any prescribed electric charge $q$ and rest mass $m$ realized as the self electric energy of the point charge, the effective radius $a$ and the nonlinear 
perturbation coupling parameter $\gamma$ are related through the equation \eq{4.20} where $H_\gamma$ is a normalized energy independent of $q$ and $a$ but depending only
on $\gamma$ and the structure of the nonlinearity function $U$ in the model. Moreover, if the normalized energy vanishes in the strong nonlinearity strength limit $\gamma\to\infty$,
which is always ensured in the finite truncation situation $\eta_0<\infty$ characterized by non-divergence in the Maxwellian limit $\gamma\to0$, and also valid in the absence of a
finite truncation threshold situation, characterized by divergence in the Maxwellian limit, in contrast, provided that the integrability condition \eq{4.19} holds which may be realized by the two growth scenarios (i) and (ii) stated in Section \ref{s3},
the effective radius $a$ of the point charge can be made arbitrarily small in the process.
\end{theorem}

\medskip

This study  further consolidates our understanding of the electric point charge problem within the unified nonlinear framework. By expressing the total self energy in terms of the normalized dimensionless integral $H_\gamma$, we move beyond specific model-dependent successes (such as those of the Born--Infeld or polynomial models) to establish the finiteness of energy as a structural consequence of the higher-order perturbation $U$. The divergence of the energy integral as $\gamma \rightarrow 0$ in the absence of a truncation threshold ($\eta_0 = \infty$) recovers the expected ultraviolet divergence of the linear Maxwell theory, whereas the persistence of finite energy in the $\eta_0 < \infty$ case highlights a fundamental distinction in how different nonlinearities regularize the point singularity. 

The property of being able to achieve an arbitrarily small effective radius for a point charge stated in Theorem \ref{th4.1} is essential when we use the point charge to model the electron. It is interesting that such a property
is unconditionally valid when the truncation bound is finite. These features and their relations will be explored further in the next section through a study of the charge and energy
distribution pictures.

\section{Distributions of energy and free charge of a point charge}
\setcounter{equation}{0}

We first consider the self energy of a point charge contained within the sphere around the charge and of radius $R>0$, denoted by $E_\gamma(R)$. Let $\eta_1$ be $\eta$ corresponding to $r=R$ determined by the constitutive equation \eq{3.1}. That is,
$\eta_1$ satisfies
\be\lb{x5.1}
\frac{a^4}{R^4}=\left(1+\gamma U'\left(\frac {\eta_1}2\right)\right)^2 \eta_1,\quad \eta_1=\beta (E^2_r)|_{r=R}.
\ee
Then we have $0<\eta_1<\eta_0$ such that
\be
    \frac{E_\gamma(R)}{4\pi}=\int_0^R {\cal H}r^2\,\dd r\\
=\frac{q^2}{a}\int^{\eta_0}_{\eta_1}h_\gamma(\eta)\,\dd\eta\equiv \frac{q^2}a H_\gamma(R),\label{5.2}
\ee
where $h_\gamma(\eta)$ is defined by \eq{4.14}.

We first consider the situation when $\eta_0<\infty$.

We see that  \eq{x5.1} defines $\eta_1=\eta_1(\gamma)$ as the inverse function of
\be\lb{x5.3}
\gamma=\gamma(\eta_1)=\frac{\left(\frac aR\right)^2 -\sqrt{\eta_1}}{\sqrt{\eta_1}U'\left(\frac{\eta_1}2\right)},
\ee
which enbles us to arrive at the limits
\begin{align}
   \lim_{\gamma\to0} \eta_1(\gamma)=&\left(\frac{a}{R}\right)^4,\quad \eta_0\geq \left(\frac aR\right)^4;\lb{5.4}\\ 
 \lim_{\gamma\to0} \eta_1(\gamma)=&
\eta_0,\quad  \eta_0<\left(\frac aR\right)^4.\lb{5.5}
\end{align}

In view of \eq{4.14}, \eq{5.2},  \eq{5.4}, and \eq{5.5}, we have
\be
    \lim_{\gamma\to0}H_\gamma(R)=\lim_{\gamma\to0}\int^{\eta_0}_{\eta_1(\gamma)}\frac{{\rm{d}}\eta}{8\eta^\frac34}=\left\{\begin{array}{lr}\frac12\left(1-\frac {R_0}R\right)\eta_0^\frac14,& R\geq R_0,\\ 0,&R<R_0,\end{array}\right.
    \ee
where
\be\lb{5.7}
R_0= a\,\eta_0^{-\frac14}
\ee
serves as an interesting {\em cutoff} radius, which is proportional to the effective radius, and renders the property that, in the $\gamma\to0$ limit, the point charge is energetically
invisible locally within the distance $R_0$.

In the limit $\gamma\to\infty$, since \eq{4.18} holds, we see that the point charge is energetically invisible globally too.

 Next consider $\eta_0=\infty$.  Since there is no finite truncation threshold, we have
\be\lb{5.8}
\lim_{\gamma\to0}\eta_1(\gamma)=\left(\frac aR\right)^4, \quad \forall\, R>0,
\ee
in view of  \eq{x5.3}.

Thus, using Fatou's lemma in \eq{5.2}, we have
    \begin{align}
        \liminf_{\gamma\to0}H_\gamma(R)\geq \int^\infty_0 \liminf_{\gamma\to0}\left(\chi_{(\eta_1(\gamma),\infty)}(\eta)h_\gamma(\eta)\right)\,\dd\eta=\int^\infty_{\left(\frac aR\right)^4}\frac{{\rm{d}}\eta}{8\eta^\frac34}=\infty, \quad\forall\, R,
    \end{align}
where $\chi_{A}$ denotes the characteristic function for the set $A$. This divergent result indicates that the limit $\gamma\to0$ energetically recovers the Maxwell theory prediction regarding a point
charge in the situation of the absence of a finite truncation bound.

When $\gamma\to\infty$, we see that \eq{4.18} is valid under the condition \eq{4.19}. 

We now consider the spatial distribution of the free charge generated by the electric field $\bf E$ of a point charge.
For this purpose, recall that the (normalized) free electric charge density as a function of $r=|{\bf x}|>0$ is given by
\be\lb{5.10}
    \rho_{\rm{free},\gamma}=\frac1{4\pi}\nabla\cdot\textbf{E}=\frac1{4\pi r^2}\frac{{\rm{d}}}{\rm{d}r}(r^2E_r),
\ee
such that, integrating it over a ball of radius $R>0$ around the point charge $q$ represented by the electric displacement field $\bf D$ as stated in \eq{a3.11} 
for all ${\bf x}\neq{\bf0}$ gives us the free charge contained in the ball to be
\begin{align}
    q_{\rm{free},\gamma}(R)=\int_{|\textbf{x}|\le R}\rho_{\rm{free},\gamma}({\bf x})\,{\rm{d}}\textbf{x}= (r^2 E_r)_{r=R}.\label{5.11}
\end{align}
Thus, in view of the constitutive equation \eq{x5.1} and \eq{5.11} we have
\begin{align}\lb{5.12}
    q_{{\rm free},\gamma}(R)=\frac{q}{a^2} R^2\sqrt{\eta_1(\gamma)}. 
\end{align}

We again have two situations to consider.

First, if $\eta_0$ is finite, then \eq{5.4} and \eq{5.5} are valid.  Thus, using \eq{5.4} and \eq{5.5} in  \eq{5.12}, we have
\be\lb{5.13}
 \lim_{\gamma\to0} q_{\rm{free},\gamma}(R)=q\left(\frac R{R_0}\right)^2,\quad R<R_0;\quad \lim_{\gamma\to0}q_{\rm{free},\gamma}(R)=q,\quad R\geq R_0.
\ee

In the limit $\gamma\to\infty$, consistency in \eq{x5.1} leads to
\be\lb{5.14}
\lim_{\gamma\to\infty} \eta_1(\gamma)=0.
\ee
In fact, differentiating \eq{x5.1} with respect to $\gamma$, we get
\be
\eta_1'\left(1+\gamma\left(U'\left(\frac{\eta_1}2\right)+\eta_1 U''\left(\frac{\eta_1}2\right)\right)\right)+2\eta_1 U'\left(\frac{\eta_1}2\right)=0,
\ee
after eliminating the common factor $1+\gamma U'\left(\frac{\eta_1}2\right)$.
Hence, in view of the condition \eq{b3.5}, we have $\eta_1'(\gamma)\leq0$. That is, $\eta_1(\gamma)$ decreases in $\gamma>0$. In view of this fact and \eq{x5.1}, it is clear that
\eq{5.14} is true.

Using this in \eq{5.12}, we have $q_{\rm{free},\gamma}(R)\to0$ as $\gamma\to\infty$ for any $R>0$.

Next, if $\eta_0=\infty$, then \eq{5.8} is valid. Inserting this into \eq{5.12}, we have
\be
\lim_{\gamma\to0} q_{\rm{free},\gamma}(R)=q,\quad\forall\, R>0.
\ee
This result is in an elegant comparison with the energy counterpart of the problem where we have shown that there is energy divergence in any ball $\{|{\bf x}|\leq R\}$ around the
point charge, in the Maxwell theory limit $\gamma\to0$.

We note that an important by-product of this study is that the free charge in a ball of any radius $R>0$ around the point charge decreases with respect to the coupling parameter $\gamma$
and vanishes as $\gamma\to\infty$ as a result of \eq{5.12} and \eq{5.14}.

Note also that, since we have already established the fact that $E_r$ is asymptotically Coulomb, we have
\be\lb{5.17}
\eta_1(\gamma)\approx \frac{a^4}{R^4},\quad R\gg1.
\ee
Inserting \eq{5.17} into \eq{5.12}, we get 
\be
\lim_{R\to\infty} q_{\rm{free},\gamma}(R)=q,
\ee
for any $\gamma$, which is consistent with the statement of Theorem \ref{th3.1}, as anticipated.

Since $q>0$, it is natural to expect that the normalized free charge density given in \eq{5.10} stays positive too for $r>0$. To examine this property, we use the relation $\eta=\eta(r)=\beta E_r^2$ to get
\be\lb{5.19}
\frac{\dd E_r}{\dd r}=\frac{\eta'(r)}{2\beta E_r}=-\frac{2q}{r^3 \left(1+\gamma\left(U'\left(\frac\eta2\right)+\eta U''\left(\frac\eta2\right)\right)\right)},
\ee
where we have inserted $\eta'(r)$ obtained by differentiating \eq{3.1} to get the equation
\be\lb{5.20}
-\frac{4a^4}{r^5}=\eta'(r)\left(1+\gamma U'\left(\frac\eta2\right)\right)\left(1+\gamma\left(U'\left(\frac\eta2\right)+\eta U''\left(\frac\eta2\right)\right)\right).
\ee
Substituting \eq{5.19} into \eq{5.10}, we have
\be
\rho_{\rm{free},\gamma}=\frac{q\gamma\eta\, U''\left(\frac\eta2\right)}{2\pi  r^3\left(1+\gamma U'\left(\frac\eta2\right)\right)\left(1+\gamma\left(U'\left(\frac\eta2\right)+\eta U''\left(\frac\eta2\right)\right)\right)},
\ee
which renders us the property $\rho_{\rm{free},\gamma}>0$ when $U''(\tau)>0$ for $\tau>0$.

We summarize our results of the study of this section regarding the distributions of self energy and free charge of a point charge as follows:

\begin{theorem}\lb{th5.1}

Consider an electric point charge in the nonlinearly perturbed Maxwell theory governed by the Lagrangian action density \eq{1.5}. The free electric charge and self energy distributions in the space
enjoy the following properties with respect to whether the model possesses a finite truncation threshold $\eta_0$ defined by \eq{x4.4}.

\begin{enumerate}
\item[(i)] For any $R>0$, the free charge, $q_{\rm{free},\gamma}(R)$,  contained in the ball of radius $R$ around the prescribed point charge $q>0$ decreases with respect to the nonlinear strength coupling parameter $\gamma$ and 
eventually vanishes as $\gamma\to\infty$. The behavior of the charge $q_{\rm{free},\gamma}(R)$ as $\gamma\to0$, on the other hand, depends on whether $\eta_0$ is finite ---

\begin{itemize}

\item If $\eta_0<\infty$, then a rescaled cutoff radius $R_0=a\eta_0^{-\frac14}$, where $a$ is the effective radius of the point charge, appears such that $q_{\rm{free},\gamma}(R)$ 
enjoys the limit property \eq{5.13} which characterizes $R_0$ as a truly effective radius of the point charge so that the ball of radius $R_0$ captures all available charge $q$ in
the zero nonlinearity limit.

\item If $\eta_0=\infty$, we have $q_{\rm{free},\gamma}(R)\to q$ as $\gamma\to0$ for any $R>0$ such that
there is no cutoff radius and the effective radius does not exhibit itself as in the finite $\eta_0$ situation. 

\item In all situations and under the condition $U''(\tau)>0$ ($\tau>0$), the free charge density is positive such that $q_{\rm{free},\gamma}(R)$ increases with respect to $R$ for any
fixed $\gamma>0$ to its limit $q$.

\end{itemize}

\item[(ii)] Similarly, the self energy of the point charge contained in the ball of radius $R$ around the point charge, $E_\gamma (R)$, as a function of $R$ and $\gamma$, has the following
properties ---

\begin{itemize}

\item If $\eta_0<\infty$, the presence of the cutoff radius $R_0$ exhibits itself through the limiting behavior
\be
\lim_{\gamma\to0}E_\gamma(R)=0,\quad R<R_0;\quad \lim_{\gamma\to0}E_\gamma(R)=\left(\frac {q^2}a\right)2\pi \left(1-\frac{R_0}R\right)\eta_0^{\frac14},\quad R\geq R_0,
\ee
which implicates that, in the Maxwell theory limit, the point charge is energetically undetectable with regard to its self energy and it becomes visible only beyond $R_0$.
Furthermore, $E_\gamma(R)$ vanishes for any $R>0$ in the limit $\gamma\to\infty$, indicating the energetic invisibility of the point charge in the strong nonlinearity or zero effective
radius situation.

\item If $\eta_0=\infty$, we have the self energy divergence $E_{\gamma}(R)\to\infty$ as $\gamma\to0$ for any $R>0$, showing a recovery of the Maxwell theory in describing the
energy of a point charge.

\end{itemize}

Thus, the presence of a finite truncation threshold $\eta_0$ is crucial for a complete and precise characterization of a point charge such that it makes the effective radius truly effective
electrically and energetically in the Maxwell theory limit $\gamma\to0$ and makes the point charge undetectable electrically and energetically in the strong nonlinear perturbation limit 
$\gamma\to\infty$ which is equivalent to the zero effective radius limit.
\end{enumerate}

\end{theorem}

\begin{figure}[h]
\centering
\begin{tikzpicture}
\begin{axis}[
    axis lines = middle,
    xlabel = $R$,
    xlabel style = {anchor=west,at={(1.02,0.08)}}, 
    ylabel = {$q_{\rm{free},0}(R)$},
    ylabel style = {anchor=south, at={(0.05,1.03)}},
    xmin = -0.1, xmax = 1.5,
    ymin = -0.12, ymax = 1.2,
    xtick = \empty,
    ytick = \empty,
    axis line style = thick,
    tick style = {line width=0.8pt},
    label style = {font=\small},
    grid = none,
    samples = 200,
    domain = 0:5,
    enlarge x limits = false,
    enlarge y limits = false,
    clip = true
]

\addplot[domain=0:1, thick, blue] {x^2};

\addplot[domain=1:5, thick, red] {1};

\addplot[mark=*, only marks, black] coordinates {(1,0)};

\addplot[mark=*, only marks, mark options={fill=white, draw=red}, red] coordinates {(1,0)};

\node at (axis cs:1,-0.02) [anchor=north] {$R_0$};

\draw[dotted](0,1)--(1,1);

\node at (axis cs:-0.095,1) [anchor=west] {$q$};

\end{axis}
\end{tikzpicture}
\caption{The behavior of the free charge distribution of a point charge in the Maxwellian limit $q_{\rm{free},0}(R)=\lim_{\gamma\to0} q_{\rm{free},\gamma}(R)$ in the ball centered around the charge and
of radius $R$. There is a cutoff radius $R_0$ resembling the effective radius such that the limiting free charge increases from zero to
its full charge $q$ only when $R$ reaches the level $R_0$. }
\label{F1}
\end{figure}
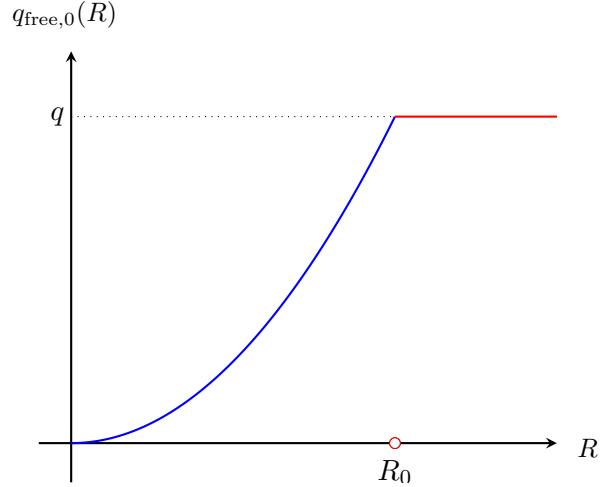

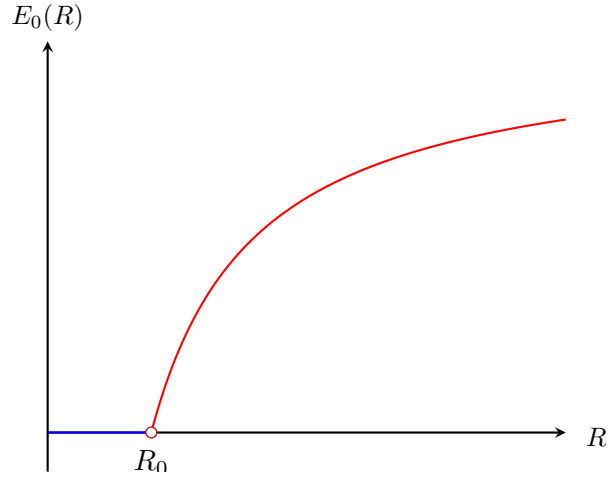
\begin{figure}[h]
\centering
\begin{tikzpicture}
\begin{axis}[
    axis lines = middle,
    xlabel = $R$,
    xlabel style = {anchor=west,at={(1.02,0.08)}}, 
    ylabel = {$E_0(R)$},
    ylabel style = {anchor=south, at={(0,1)}},
    xmin = 0, xmax = 5,
    ymin = -0.1, ymax = 1.0,
    xtick = \empty,
    ytick = \empty,
    axis line style = thick,
    tick style = {line width=0.8pt},
    label style = {font=\small},
    grid = none,
    samples = 200,
    domain = 0:5,
    enlarge x limits = false,
    enlarge y limits = false,
    clip = true
]

\addplot[domain=0:1, thick, blue] {0};

\addplot[domain=1:5, thick, red] {1 - 1/x};

\addplot[mark=*, only marks, black] coordinates {(1,0)};

\addplot[mark=*, only marks, mark options={fill=white, draw=red}, red] coordinates {(1,0)};

\node at (axis cs:1,-0.02) [anchor=north] {$R_0$};

\end{axis}
\end{tikzpicture}
\caption{The behavior of the self energy distribution of a point charge in the Maxwellian limit $E_0(R)=\lim_{\gamma\to0} E_\gamma(R)$ in the ball centered around the charge and
of radius $R$. There is a cutoff radius $R_0$ resembling the effective radius such that the limiting energy vanishes below $R_0$ and becomes measurable above $R_0$. Thus
the point charge is locally energetically undetectable even in the Maxwell theory limit as a consequence of a nonlinear perturbation with a finite truncation threshold.}
\label{F2}
\end{figure}

Figures \ref{F1} and \ref{F2} are illustrative plots of the behaviors of the free charge and self energy distributions of a point charge contained in a ball centered about the charge, respectively, in the
Maxwell theory limit $\gamma\to0$ when the nonlinear perturbation possesses a finite truncation threshold. These plots show clearly the effects of a cutoff radius $R_0$ associated with
the effective radius of the point charge. In particular, it is seen that there is a sharp disparity between charge and energy below $R_0$ and that such a disparity disappears in the full-space limit.

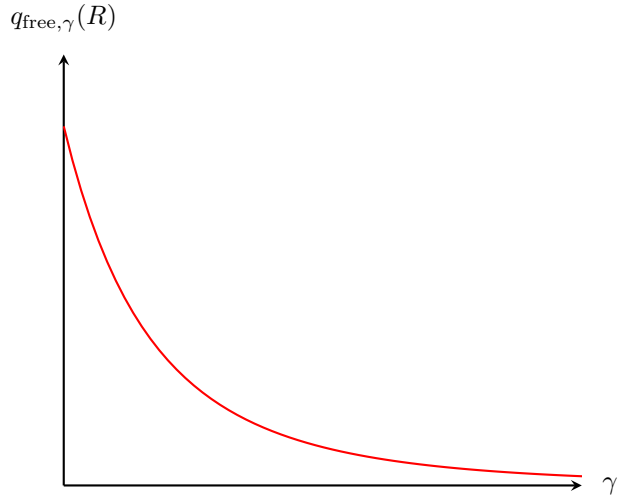
\begin{figure}[h]
\centering
\begin{tikzpicture}
\begin{axis}[
    axis lines = middle,
    xlabel = $\gamma$,
    xlabel style = {anchor=west,at={(1.02,0)}}, 
    ylabel = {$q_{\rm{free},\gamma}(R)$},
    ylabel style = {anchor=south, at={(0,1.03)}},
    xmin = 0, xmax = 1.5,
    ymin = -0, ymax = 1.2,
    xtick = \empty,
    ytick = \empty,
    axis line style = thick,
    tick style = {line width=0.8pt},
    label style = {font=\small},
    grid = none,
    samples = 200,
    domain = 0:5,
    enlarge x limits = false,
    enlarge y limits = false,
    clip = true
]


\addplot[domain=0:5, thick, red] {1/(1+x)^4};




\end{axis}
\end{tikzpicture}
\caption{The behavior of the free charge $q_{\rm{free},\gamma}(R)$ contained in the ball centered around the charge and
of any fixed radius $R>0$ as a monotone decreasing function of the nonlinear coupling parameter $\gamma$ and vanishing at $\gamma=\infty$, in all situations, whether or not there is
a finite truncation threshold. The same vanishing property is also valid for the self energy in general. Thus the point charge is electrically and energetically undetectable in the strong nonlinearity limit as a universal property of the theory.}
\label{F3}
\end{figure}

Figure \ref{F3} shows the free charge contained in a ball as a function of the parameter $\gamma$ as described in Theorem \ref{th5.1}.

\medskip

The results of Theorem 5.1 provide the quantitative foundation for the  most significant phenomenological features of the nonlinearly perturbed Maxwell theory \eq{1.5} ---  the technical transition between classical Maxwellian behavior and the local ``invisibility" of the point charge. By distinguishing between models with and without a finite truncation threshold, the theorem elucidates how the nonlinear perturbation governs the distribution of charge and energy. In the presence of a truncation threshold, the existence of a rescaled cutoff radius allows for a refined characterization of the charge and energy distributions as they approach the Maxwellian limit. Conversely, in the absence of such a threshold, the theory naturally recovers the standard Maxwellian point-charge description. Most crucially, the theorem establishes that in the strong nonlinearity limit, both the charge and self-energy within any finite ball around the point source vanish. This universal property provides a rigorous field-theoretic basis for the local electrical and energetical invisibility of the electron, reconciling the existence of a finite-energy point particle with its lack of detectable structure at small scales.

In the next section, we consider some examples as concrete realizations of our general theory.

\section{Some concrete models realizing the existence and absence of finite truncation thresholds}\label{sec6}
\setcounter{equation}{0}

Our general theory is developed out of our recent study \cite{LY2} based on a Born--Infeld theory type nonlinearity  where there is a finite truncation threshold $\eta_0$  and a polynomial theory formalism where
a finite truncation threshold $\eta_0$
is absent. In this section, we consider some other models along these two lines.

Let $\eta_0>0$ be a fixed constant. The following examples of nonlinearity functions are the realizations of a perturbed Maxwell theory with $\eta_0$ as its
finite truncation threshold and enjoying all the properties obtained in the previous sections:
\bea
U\left(\frac\eta2\right)&=&\ln\eta_0^k-\ln(\eta_0^k-\eta^k),\quad 0\leq\eta<\eta_0,\\
U\left(\frac\eta2\right)&=&\frac1{\eta_0^k-\eta^k}-\frac1{\eta_0^k},\quad 0\leq\eta<\eta_0,
\eea
where $k=2,3,\dots$, and
\bea
U\left(\frac\eta2\right)&=&\eta\left(\eta_0-\sqrt{\eta_0-\eta}\right),\\
U\left(\frac\eta2\right)&=&\eta\tan\left(\frac{\pi\eta}{2\eta_0}\right),
\eea

Moreover, if we set
\bea
U\left(\tau\right)&=&{\rm{e}}^{\tau^k}-1,\quad k=2,3,\dots,\\
U(\tau)&=&\tau (\e^\tau-1),\\
U(\tau)&=&\tau \arctan \tau,\\
U(\tau)&=&\left(\sinh (\tau)\right)^k,\quad k=2,3,\dots,\\
U(\tau)&=&\tau\,\tanh (\tau),\lb{6.8}
\eea
we obtain some models where a finite truncation threshold is absent and our theory in this context is applicable. A difference of the model \eq{6.8} is that although the
condition $U''(\tau)\geq0$ is not satisfied but the weaker condition $U'(\tau)+2\tau U''(\tau)\geq0$ is valid which ensures all the required structures of the theory are available.


\medskip

The specific families of nonlinearities discussed above serve as concrete realizations of the general formalism established in this work. From a phenomenological perspective, these models provide a flexible framework for investigating the internal electronic and energetic structures of point charges. By adjusting the high-order perturbation terms, one can systematically explore the transition between the classical macroscopic Maxwell regime and the micro-scale regime where the effective radius of the charge becomes arbitrarily small, consistent with experimental bounds on the electron size \cite{Deh1,Deh2}.

Beyond the scope of flat-space electrodynamics, these nonlinearly perturbed theories offer significant utility in curved spacetime contexts. Specifically, they provide the necessary matter source for the construction of regularity-enhanced  charged black hole solutions. In such gravitational settings, the nonlinearities prevent the divergence of the electric field at the origin, thereby regularizing the spacetime metric. Furthermore, such nonlinear field theories find natural applications in k-essence cosmology, where the non-canonical kinetic terms associated with the electromagnetic field or its scalar analogs can drive late-time cosmic acceleration or model dark energy dynamics \cite{Yang1}. These applications underscore the robustness of the universal properties identified herein, suggesting that the exclusion of monopoles and the finite-energy nature of point charges offer fundamental features across diverse physical landscapes.


\section{Conclusions and comments}

In this work, we have examined a general class of nonlinearly perturbed Maxwell theories, characterized by a Lagrangian action density in which the familiar Maxwell term is augmented by a higher-order nonlinear correction controlled by a coupling parameter $\gamma$ and a scale parameter $\beta$. The nonlinearity function $U$ is assumed only to satisfy the
characteristic conditions --- $U(0)=0$ and $U'(0)=0$ --- ensuring that the Maxwell theory is recovered in the weak-field limit. Despite this generality, the model exhibits several universal properties that are largely independent of the specific form of $U$.

\subsection*{Universal exclusion of magnetically charged finite-energy configurations}

Perhaps the most robust conclusion is that, across all admissible choices of $U$, the theory does not support finite-energy, point-like magnetic monopoles or dyons. This result follows from the inevitable singular behavior of the magnetic field near a monopole source, which forces the energy density to diverge in a manner that cannot be tamed by any nonlinear modification of the form considered here. Thus, while electric point charges are permitted (and indeed, can carry finite energy), magnetic counterparts are fundamentally excluded --- a sharp asymmetry between electricity and magnetism that persists even in the most general setting.

\subsection*{Two distinct regimes for electric point charges}

The behavior of electric point charges falls naturally into two families, distinguished by whether the nonlinearity imposes a finite upper bound on the electrostatic field strength --- a \emph{truncation threshold} $\eta_0$.

\begin{itemize}
\item \textbf{Finite-threshold models} (as exemplified by the Born--Infeld type and the logarithmic or rational examples in Section \ref{sec6}) exhibit a built-in maximal field. In these theories, the Maxwell limit $\gamma \to 0$ is smooth: the self-energy remains finite, and the effective radius $a$ can be driven arbitrarily small by taking the nonlinear coupling $\gamma$ sufficiently large. Moreover, in this limit, the point charge becomes both electrically and energetically invisible inside a sphere of radius $R_0 = a\eta_0^{-1/4}$ --- a genuine \emph{cutoff radius} that renders the charge locally undetectable. This offers a classical mechanism for why the electron, as a point particle, has never been observed experimentally.

\item \textbf{Infinite-threshold models} (such as polynomial or exponential nonlinearities) lack such a field bound. In the Maxwell limit $\gamma \to 0$, the self-energy diverges of an 
electric point charge, recovering the familiar ultraviolet catastrophe of linear electromagnetism. However, for any finite $\gamma > 0$, the energy remains finite, and in the strong-coupling limit $\gamma \to \infty$, the effective radius again shrinks to zero, and the charge becomes globally invisible. Thus, even without a truncation threshold, the theory still accommodates an arbitrarily small effective radius, albeit with a different limiting behavior as $\gamma \to 0$.
\end{itemize}

\subsection*{Arbitrary smallness of the effective radius as a universal phenomenon}

A central message of this paper is that the ability to make the effective radius of an electric point charge arbitrarily small --- and thereby render the charge locally undetectable --- is a \emph{universal} feature of the general model, regardless of whether a finite truncation threshold exists. The difference lies only in the manner in which this limit is approached and in the behavior of the Maxwell limit. This finding suggests that classical nonlinear electrodynamics, properly formulated, can accommodate a point-like electron whose spatial extent is effectively zero for all practical purposes, without incurring infinite energy.

\subsection*{The role of the truncation threshold}

While not necessary for the existence of an arbitrarily small effective radius, the presence of a finite truncation threshold $\eta_0$ does confer additional structure. In particular, it gives rise to a sharp cutoff radius $R_0$ in the zero-nonlinearity limit, below which the point charge is completely invisible --- a kind of classical ``screening'' that is absent in infinite-threshold models. This cutoff is not merely a mathematical artifact; it directly controls the onset of charge and energy detection as one moves outward from the source.

\subsection*{Comparison with previous work and experimental context}

Our analysis unifies and extends earlier studies of the Born--Infeld theory and its polynomial relatives. The Born--Infeld model, while historically important, predicts an effective electron radius on the order of $10^{-13}$~cm --- far larger than current experimental bounds (below $10^{-20}$~cm). By contrast, the general framework presented here shows that, by taking $\gamma$ sufficiently large (i.e., by entering the strongly nonlinear regime), the effective radius can be made arbitrarily small, easily satisfying all observational constraints. Thus, nonlinear electrodynamics remains a viable classical paradigm for the electron, free from the inconsistencies that plagued the original Maxwell theory and the Born--Infeld type theories.

\subsection*{Outlook and open questions}

Several directions invite further investigation. First, the dynamical behavior of these theories --- particularly the propagation of waves and the scattering of charges --- has not been explored.  Second, the quantization of such nonlinear field theories, while notoriously difficult, could shed light on whether the classical invisibility of the point charge survives quantum corrections. Third, the interplay between the nonlinear electromagnetic structure and gravity (for example, that exhibited in black hole solutions and in the coupled quintessence and k-essence cosmology) may reveal new physics, especially in strong nonlinearity regime corresponding to the effective radius being extremely small. Finally, the possibility of experimentally distinguishing between finite-threshold and infinite-threshold models, perhaps through precision measurements of the electron's anomalous magnetic moment or through high-field tests of quantum electrodynamics, remains an open and intriguing question.

\subsection*{Closing remark}

In summary, this work demonstrates that a general, minimally modified Maxwell theory --- with a higher-order nonlinear perturbation --- exhibits a rich and largely universal set of properties regarding point charges: finiteness of energy, exclusion of magnetic monopoles, and the capacity to render a point charge arbitrarily small and locally undetectable. These features suggest that classical nonlinear electrodynamics, far from being a mere historical curiosity, is able to offer a consistent and flexible framework for modeling the electron as a point particle.

\subsection*{Note on the causality conditions}

Note that the Pleba\'{n}ski class \cite{Ple} of the generalized electrodynamical Lagrangian $\cal L$ takes the form
${\cal L}={\cal L}(F,G)$ where
\be
F=\frac12 F_{\mu\nu}F^{\mu\nu}={\bf B}^2-{\bf E}^2,\quad G=\frac14 F_{\mu\nu}\tilde{F}^{\mu\nu}={\bf B}\cdot{\bf E},
\ee
which contains our theory \eq{1.5} as a special case since \eq{1.5} now reads
\be\lb{7.2}
{\cal L}=-\frac12 F +\frac\gamma\beta U\left(-\frac\beta2 F+\frac{\beta\kappa^2}2 G^2\right).
\ee
In \cite{Sche} (see also \cite{RT}), subject to the condition ${\cal L}_F<0$, the conditions that ensure causality of the theory of
the Pleba\'{n}ski class are found to be
\bea
{\cal L}_{FF} {\cal L}_{GG}&\geq& {\cal L}^2_{FG},\lb{7.3}\\
{\cal L}_{FF}&\geq&0,\quad {\cal L}_{GG}\geq0,\lb{7.4}\\
F\left(\frac14 {\cal L}_{GG}-{\cal L}_{FF}\right)&>&{\cal L}_F+2G{\cal L}_{FG}+\left({\cal L}_{FF}+\frac14{\cal L}_{GG}\right)\sqrt{F^2+4G^2}.\lb{7.5}
\eea
For our model \eq{7.2}, it is clear that \eq{7.3} and \eq{7.4} impose the conditions $U'\geq0$ and $U''\geq0$, which ensure the validity of our 
condition \eq{b3.5}. Moreover, in view of our exclusion principle of monopole and dyons as stated in Theorem \ref{th2.1}, the finite-energy condition
dictates ${\bf B}={\bf 0}$ or $G=0$. (It should be emphasized that such an exclusion principle is not available in the Born theory \cite{B1,B2} or in the Born--Infeld theory \cite{BI1,BI2}.) Inserting this condition into \eq{7.5} and using \eq{7.2}, we see that \eq{7.5} becomes
\be
1+\gamma U'(\tau)(1-\kappa^2 {\bf E}^2)>0,\quad \tau=\frac12\beta {\bf E}^2,
\ee
which is ensured by imposing the sufficient condition
\be\lb{7.7}
\kappa^2{\bf E}^2\leq 1.
\ee
In the context of the study here, when a finite truncation threshold $\eta_0>0$ is present, then we have $\beta{\bf E}^2\leq \eta_0$ such that \eq{7.7}
reads
\be\lb{7.8}
\kappa^2\leq\frac\beta{\eta_0},
\ee
which spells out a bound for the electromagnetic interaction strength even though there is no presence of a magnetic field due to the finite-energy
condition for localized charges. This result is surprising. Interestingly, for the Born--Infeld type nonlinearity \cite{LY2}
\be
U(\tau)=1-\sqrt{1-(2\tau)^k},\quad k\geq2,
\ee
we have $\eta_0=1$. Thus \eq{7.8} yields the bound $\kappa^2\leq\beta$. In particular, in the classical Born parameter situation \cite{B1,B2,BI1,BI2}, we have $\kappa^2=\beta$
such that the bound \eq{7.8} is automatically satisfied for all $k\geq2$.

Other examples such as those presented in Section \ref{sec6} and elsewhere satisfying the causality conditions can be worked out similarly.

In the case of a continuously distributed magnetic source, the finite-energy condition always holds. In this situation, if the electric field is absent, it can be
shown that \eq{7.3}, \eq{7.4}, and \eq{7.5} are ensured by $U'\geq0$, $U''\geq0$, and the second condition in \eq{b3.5}, namely,
\be
1+\gamma U'(\tau)+2\gamma\tau U''(\tau)>0.
\ee

{\fontsize{10}{12}\selectfont

}
\end{document}